\documentclass{aa} 

\usepackage{graphicx}
\usepackage{subcaption}
\usepackage{natbib}
\usepackage{comment}
\usepackage{array}
\usepackage{xcolor}
\usepackage{catchfile}
\usepackage{txfonts}
\usepackage{hyperref}
\hypersetup{
    colorlinks=true,
    linkcolor=blue,
    citecolor=blue
    }
\begin{document}

   \title{Dynamic atmosphere and wind models of C-type asymptotic giant branch stars}

   \subtitle{Influences of dust optical data on mass loss and observables}

   \author{E. Siderud
          \and
          K. Eriksson
          \and
          S. Höfner
          \and
          S. Bladh
          }

   \institute{Theoretical Astrophysics, Division for Astronomy and Space Physics, Department of Physics and Astronomy, Uppsala University,
              Box 516, 751 20 Uppsala, Sweden\\
              \email{emelie.siderud@physics.uu.se}
             }

   \date{Received 2 August 2023 / Accepted 24 January 2025}
 
  \abstract
   {Mass loss through stellar winds governs the evolution of stars on the asymptotic giant branch (AGB). In the case of carbon-rich AGB stars, the wind is believed to be driven by radiation pressure on amorphous carbon (amC) dust forming in the atmosphere. The structural complexity of amC is evident from the diversity of laboratory optical data that are available in the literature. Consequently, the choice of dust optical data will have a significant impact on atmosphere and wind models of AGB stars.}
   {We compare two commonly used optical data sets of amC and investigate how the wind characteristics and photometric properties resulting from dynamical models of carbon-rich AGB stars are influenced by the micro-physical properties of dust grains. }
   {We computed two extensive grids of carbon star atmosphere and wind models with the DARWIN 1D radiation-hydrodynamical code. A defining feature of these models is a self-regulating feedback between the time-dependent dynamics, grain growth, and dust optical properties. Thus, they are  able to predict combinations of mass-loss rates, wind velocities, and grain sizes for given stellar parameters and micro-physical data. Each of the two grids uses a different amC optical data set. The stellar parameters of the models were varied in terms of the effective temperature, luminosity, stellar mass, carbon excess, and pulsation amplitude to cover a wide range of possible combinations. A posteriori radiative transfer calculations were performed for a sub-set of the models, resulting in photometric fluxes and colours.}
   {We find small, but systematic differences in the predicted mass-loss rates for the two grids. The grain sizes and photometric properties are affected by the different dust optical data sets. Higher absorption efficiency leads to the formation of a greater number of grains, which are smaller. Models that are obscured by dust exhibit differences in terms of the covered colour range compared to observations, depending on the dust optical data used.}
   {An important motivation for this study was to investigate how strongly the predicted mass-loss rates depend on the choice of dust optical data, as these mass-loss values are more frequently used in stellar evolution models. Based on the current results, we conclude that mass-loss rates may typically differ by about a factor of two for DARWIN models of C-type AGB stars for commonly used dust optical data sets.}


   \maketitle
%
%-------------------------------------------------------------------

\section{Introduction}

A major fraction of all stars (i.e. those with initial masses of about $0.8-8\,M_\odot$) will evolve into asymptotic giant branch (AGB) stars. These AGB stars are characterised by substantial mass loss through stellar winds. During the late stages of the AGB phase, the star is powered by alternating burning of H and He in thin shells surrounding an inert carbon-oxygen core. This cyclic process is connected to a phenomenon called a He-shell flash (or a thermal pulse), where the convective zone in the outer envelope of the star reaches down to the region between the two shells and dredges up carbon-rich material to the surface (see e.g. \citealt{herwig2005} for details). The gradual dredge-up of freshly produced carbon will alter the atmospheric abundances and may turn an M-type AGB star into a carbon star (C-type) with a carbon-to-oxygen ratio exceeding unity (C/O $>1$).

The majority of AGB stars are pulsating long-period variables and their atmospheres provide favourable conditions for molecules and dust grains to form. The pulsations generate shock-waves in the atmosphere that push material to regions where the temperatures are cool enough for dust to condense. Radiation pressure exerted by the luminous star causes the newly formed dust particles to accelerate outwards and transfer momentum to the surrounding gas by collisions, resulting in a stellar wind. 
The theoretical understanding of the wind mechanism has evolved with the progress of dynamical models, starting with 1D steady-state outflows, followed by spherically symmetric time-dependent models, which take the effects of radial pulsation on atmospheric structure and dust formation into account (see \citealt{hofner2018} for a review). The most recent developments are 3D radiation-hydrodynamical models, covering the convective and pulsating stellar interior, as well as the dynamical atmosphere and dust-driven wind (\citealt{freytag2023}). However, since such 3D models are very computationally demanding, 1D models are still used for studying effects of physical processes (e.g. drift between gas and dust, \citealt{sandin2020}) and for computing extensive model grids that span wide ranges in terms of the stellar parameters, as required for application to stellar evolution models (e.g. \citealt{eriksson2023}). In this paper, it is the latter aspect that we focus on.

One particular point that all types of dynamical models, independent of their level of sophistication, have in common is that the optical properties of the dust will determine its efficiency as a wind driver. The most common type of dust present in the circumstellar envelopes of C-type AGB stars is amorphous carbon (amC). It is considered the most likely wind-driving dust species around carbon stars. The chemical structure of amC can be simplified as a random network of $\mathrm{sp^2}$ and $\mathrm{sp^3}$ hybridisation states and the $\mathrm{sp^2/sp^3}$ ratio can be used to distinguish between structures that are more graphite-like (more $\mathrm{sp^2}$ bonds) and more diamond-like (more $\mathrm{sp^3}$ bonds). Carbon dust that is synthesized in laboratories will be characterized by different $\mathrm{sp^2/sp^3}$ ratios, depending on the processes used when preparing the samples. These differences in the micro-physical structure affect the optical properties of the material. Another reason for the dissimilarities between various data in the literature is the application of different techniques to measure the optical properties of the samples.

When choosing optical data for models of dust-driven winds, it is critical to know how differences in the data affect the resulting properties of the models. Currently, there are two major types of dynamical wind models for C-rich AGB stars  used in the literature, namely:

(i) time-dependent radiation-hydrodynamical models including effects of stellar pulsation and dust grain growth, aimed at predicting mass-loss rates and other properties for given stellar parameters (e.g. \citealt{wachter2002,wachter2008,mattsson2010,eriksson2014,bladh2019a}) and 

(ii) steady-state models of dust-driven outflows including grain growth, that predict wind velocities, dust properties and observables for given stellar parameters and given mass-loss rate (e.g. \citealt{ferrarotti2006,nanni2013,nanni2014, nanni2016, nanni2019b}). 

Both types of models necessarily have a range of physical input parameters (stellar properties and micro-physical data). A meaningful analysis needs to take this into account, isolating the effects of the dust optical data by comparing models with otherwise identical basic assumptions and input parameters.
The first type of models, discussed in the present paper, predicts the dependency of mass-loss rates on stellar parameters, thereby providing critical input for stellar evolution calculations (e.g. \citealt{pastorelli2019,marigo2020}). The second type can be used to deduce dust production rates from observations of individual stars and stellar populations, by applying results of stellar evolution models (e.g. \citealt{nanni2019b}). In all cases, the outcome is affected by the choice of dust optical data. 

\citet{nanni2016} and \citet{nanni2019} tested the influence of different optical data sets for amC in the latter type of models using Mie theory to compute size-dependent dust grain opacities and comparing the resulting photometric properties to observations of AGB stars in the Magellanic Clouds. Regarding the first type of models, \citet{andersen1999} performed a detailed investigation of how different amC laboratory optical data influenced self-consistent, time-dependent models of carbon star atmospheres and winds, based on grey radiation hydrodynamics equations. \citet{andersen2003} followed up with a similar investigation for time-dependent dynamic models that include frequency-dependent radiative transfer, computed with an earlier version of the code used in the present paper. They showed that the choice of dust optical data directly influences the structure and wind properties of the models. However, \citet{andersen1999, andersen2003} applied the small particle limit approximation for computing dust optical properties and they only used a limited range of stellar parameter combinations.

In this paper, we compare two extensive grids of time-dependent atmosphere and wind models for C-type AGB stars, produced with the DARWIN code (\citealt{hofner2016}). The models use different sets of optical data from the literature (\citealt{rouleau1991} and \citealt{jager1998}), but their basic assumptions and input parameters are otherwise identical. We investigate the influence of the optical data sets on the wind and dust grain properties for the full grids, as well as on synthetic photometry for selected models. In contrast to earlier works (e.g. \citealt{andersen1999,andersen2003,mattsson2010,eriksson2014}), which relied on the small particle limit approximation for dust opacities, the new models presented here include size-dependent optical properties of dust grains, using Mie theory for spherical particles. This approach was applied in both the computation of dynamical structures and the resulting observables. 

This work is a follow-up of the recent paper by \citet{eriksson2023}. There, the effects of using grain size-dependent dust opacities, in contrast to the small particle approximation, were discussed. That study considered both wind characteristics and photometric properties for a large grid of DARWIN models, based on the optical data from \citet{rouleau1991}. While the photometric properties of the models were seen to change drastically when replacing the small particle approximation with size-dependent dust opacities, the predicted mass loss rates were not significantly affected. 

The comparative study presented here is mainly motivated by the fact, that the mass-loss rates of C-rich AGB stars predicted by radiation-hydrodynamical models are increasingly used in stellar evolution calculations (e.g. \citealt{pastorelli2019,marigo2020}). Therefore, the inherent uncertainties due to differences in optical data for amC need to be investigated.

The underlying physics and the set-up of the model grids presented here are similar to the approach of \citet{eriksson2023}; however, in this work, we  cover a wider stellar parameter range and use a second set of optical data.
In Sect. \ref{sec:method}, we give a summary of the modelling methods and a description of the different optical data sets. An overview of the physical parameters that define the model grid is also given. The dynamic and photometric properties of the models in the two grids are presented and compared in Sect. \ref{sec:resultmodel}. Furthermore, the results of the models are compared to observational data and other models from the literature in Sect. \ref{sec:resultobs}. Section \ref{sec:sum} gives a short summary and presents the main conclusions.

%--------------------------------------------------------------------

\section{Method}
\label{sec:method}

\subsection{DARWIN models}
\label{sec:methoddarwin}
The models of atmospheres and winds presented here were computed with the 1D radiation-hydrodynamics code DARWIN (Dynamic Atmosphere and Radiation-driven Wind models based on Implicit Numerics, see \citealt{hofner2016} and references therein), using frequency-dependent opacitites for the gas and dust. The models assume a spherically symmetric structure of the atmosphere and wind, with the inner boundary situated just below the photosphere. To obtain the radial structures of gas and dust, the time-dependent equations of radiation-hydrodynamics (describing the conservation of mass, momentum, and energy) were solved together with a set of equations describing the non-equilibrium condensation and evaporation of dust. The treatment of dust formation is based on a gas-kinetical approach and represents the dust properties at a given point in space and time by moments of the grain size distribution, weighted with powers of the grain radius (see \citealt{gauger1990,hofner2003} and references therein). The amC grains have been assumed to grow by addition of carbon from the gas phase through reactions with C, $\mathrm{C_2}$, $\mathrm{C_2H}$, and $\mathrm{C_2H_2}$, and to shrink by the reverse reactions, depending on the thermodynamical conditions. Micro-physical parameters inherent to such a description of dust formation (e.g. sticking probabilities) were kept constant within this study to isolate the effects of the dust optical data on wind properties. The formation of new seed particles was treated by classical nucleation theory (see \citealt{andersen2003} for a discussion of the underlying assumptions). This approach makes it unnecessary to introduce the seed particle abundance as a free parameter, as done in other recent studies (e.g. \citealt{nanni2016,nanni2019}). It means that both the abundances and sizes of dust grains are results of the models, not input parameters. The grains are assumed to remain position-coupled to the gas from which they form. The consequences of particular model assumptions are discussed in Sect. \ref{sec:modelass}.

The starting structure of each model is a hydrostatic dust-free atmosphere, defined by the current stellar mass, luminosity, effective temperature, and elemental abundances. The latter are fixed in the present study, except for the carbon excess (C$-$O), while the combinations of mass, luminosity and effective temperature values cover a range suitable for C-rich AGB stars (see Sect. \ref{sec:gridparam}). Stellar pulsations are simulated by sinusoidal variations of the radius and luminosity at the inner boundary, defined by pulsation period, velocity amplitude and luminosity amplitude (see \citealt{hofner2016}, their App. B, and below). As the pulsations are gradually ramped up, the initially compact hydrostatic atmosphere turns into an extended dynamical atmosphere that includes complex physical processes such as the propagation of shock waves or the formation and destruction of dust. If a wind develops, the outer boundary (initially located close to the photosphere) follows the outflow up to a point where it has reached its terminal velocity. Typically, this is the case around 25 stellar radii, where acceleration has dropped to about 1 percent of the value in the dust formation region or less (we note that both gravitational and radiative forces decrease as $1/r^2$). The resulting models consist of snapshots of the atmosphere and wind structures that together form long time series covering hundreds of pulsation periods. For a detailed description of the physical equations, basic assumptions and numerical methods, we refer to \citet{hofner2003,hofner2016}.

In summary, the models have two types of input parameters, namely, the micro-physical properties of gas and dust (which are kept constant here, except for the dust optical data) and the stellar parameters (including pulsation properties), which have to be specified for each model. In the current setup, this results in pairs of models where the input differs by the chosen set of optical data only. Mass-loss rates, wind velocities, dust grain abundances, and grain sizes are direct results of the simulations, which can be compared for each model pair or for the grids as a whole. 

Observables, such as photometric fluxes and colours, can be computed a posteriori, using snapshots of the structures (see Sect. \ref{sec:synth}). However, these computations are demanding to produce and analyze and we have therefore restricted them to a representative subset of models in the present study. These results are used for checking if the models fall within observed ranges and if there are significant differences between the two sets of models. They are not meant to be an exhaustive set of data that can be applied to the interpretation of observations, which is beyond the scope of this paper.

\subsection{Optical properties of dust}
Since dust opacities play a decisive role for both wind driving and observable properties, this section presents a detailed discussion of the underlying assumptions.

The models presented here include a size-dependent description of dust opacities, as used, for instance, in  \citet{eriksson2023}.
The total opacity at a wavelength, $\lambda$, of an ensemble of spherical dust grains with radii, $a_{gr}$, embedded in a gas of mass density $\rho$, can be formulated as
\begin{equation}
\label{eq:totalopac}
    \kappa_\lambda = \frac{\pi}{\rho} \int\limits_0^{\infty}a_{gr}^2\,Q(a_{gr},\lambda)\,n(a_{gr})\,da_{gr},
\end{equation}

\noindent where $n(a_{gr})\,da_{gr}$ is the number density of grains in the size interval $da_{gr}$ around $a_{gr}$, and $Q(a_{gr},\lambda)$ is the efficiency factor, defined as the radiative cross-section $C(a_{gr},\lambda)$ divided by the geometrical cross-section of the grains, namely,
\begin{equation}
    Q(a_{gr},\lambda)=\frac{C(a_{gr},\lambda)}{\pi a_{gr}^2}.
\end{equation}

 \noindent In the DARWIN models of carbon stars, the dust particles at a distance, $r,$ from the stellar center and at a time, $t,$ are described in terms of moments, $K_i(r,t),$ of the grain size distribution, weighted with a power $i$ of the grain radius,
\begin{equation}
    K_i(r,t) \propto \int\limits_0^{\infty}a_{gr}^i\,n(a_{gr})\,da_{gr} \qquad (i=0,1,2,3).
\end{equation}
\noindent From this definition it follows that $K_0$ is proportional to the total number density of grains, while $K_1$, $K_2$, and $K_3$ are related to the average radius, geometric cross-section, and volume of the grains, respectively. By defining $Q^\prime = Q/a_{gr}$, the opacity can be rewritten as
\begin{equation}
\label{eq:opacityQprime}
    \kappa_\lambda = \frac{\pi}{\rho} \int\limits_0^{\infty}Q'(a_{gr},\lambda)\,a_{gr}^3\,n(a_{gr})\,da_{gr},
\end{equation}
\noindent where the integral resembles the definition of moment $K_3$,  
apart from the factor $Q^\prime$. For particles much smaller than the relevant wavelengths, the absorption efficiency is proportional to $a_{gr}$ (making $Q^\prime$ a constant that can be taken out of the integral), and the scattering efficiency is proportional to $a_{gr}^4$ (becoming negligible for very small grains). Therefore, the small particle limit is computationally convenient and has been used in many earlier studies.
When dealing with grains that have sizes comparable to the wavelengths under consideration, 
however, the values of $Q^\prime$ for absorption and scattering are
intricately dependent on the grain radius. To determine $Q^\prime$ in such cases, numerical computations based on Mie theory are necessary. 
Furthermore, in that case the grain size distribution at each point of the atmosphere has to be known to evaluate the integral over grain sizes in Eq. \ref{eq:opacityQprime}. In principle, the grain size distribution can be reconstructed from the known time evolution of the moments $K_i$, but in practice this entails a prohibitive computational effort when dealing with large grids of models, as required in the context of stellar evolution. Therefore, when computing the dynamic models, we use an approximate treatment of the size-dependent dust opacities that relies on the available moments $K_i$, as introduced by \citet{mattsson2011}. 
Assuming that the grain sizes at each point in the model are well represented by the average grain radius $ \langle a_{gr}\rangle $, which is derived from moment $K_1$ of the size distribution, we can rephrase Eq. \ref{eq:opacityQprime} as follows:
\begin{equation}
\label{eq:opacityRe}
    \kappa_\lambda = \frac{\pi}{\rho}\,Q'(\langle a_{gr}\rangle,\lambda) \int\limits_0^{\infty}a_{gr}^3\,n(a_{gr})\,da_{gr} \propto Q'(\langle a_{gr}\rangle)\,K_3\frac{1}{\rho}.
\end{equation}

\noindent For the models described in this paper, the absorption and scattering efficiency factors, along with the mean scattering angle, are determined using Mie theory for spherical particles\footnote{Program BHMIE, originally from \citet{bohren1983}, modified by B.T. Draine  (\url{https://www.astro.princeton.edu/~draine/scattering.html})} and two different optical data sets of amC (further described below).
The radiative pressure efficiency factor, $Q_{acc}(a_{gr},\lambda)$, is a combination of true absorption and scattering, namely,
\begin{equation}
    Q_{acc} = Q_{abs} + (1-g_{sca})Q_{sca},
\end{equation}
\noindent with $g_{sca}$ as an asymmetry factor, denoting the mean over all directions of the cosine of the scattering angle where $g_{sca}=1$ corresponds to true forward scattering (see e.g. \citealt{kruegel2003}).

\subsection{Laboratory measurements of amorphous carbon}
\label{subs:lab}

\begin{figure}
\centering
      \includegraphics[clip,width=\hsize]{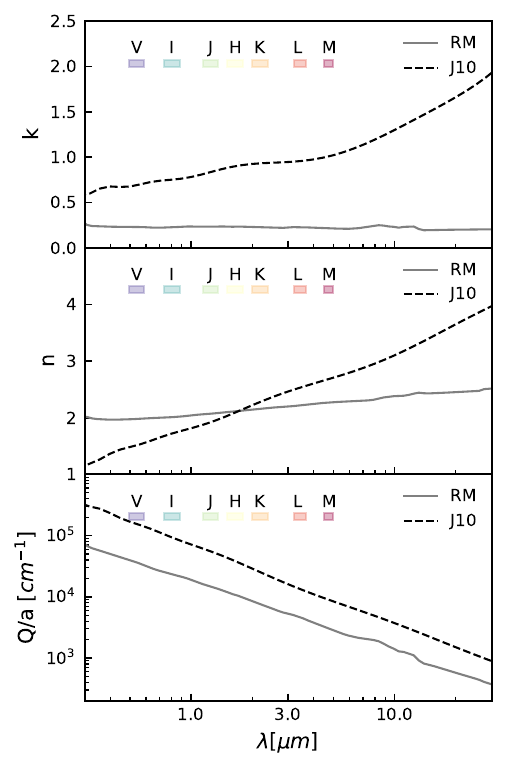}
  \caption{\textit{Top and middle panel: }Optical data, $n$ and $k$, of \citet{rouleau1991} (solid line) and \citet{jager1998} (dashed line) samples in the wavelength region $0.3-30\,\mathrm{\mu m}$. \textit{Bottom panel:} Corresponding $Q^\prime$ for the small particle limit approximation. The wavelength coverage of different photometric filters is illustrated in the figure.}
    \label{fig:nk}
\end{figure}

The optical properties of a material can be described by the complex refractive index, $m=n+ik$, where the real part, $n$, determines the phase velocity of light within the material, and the imaginary part, $k$, determines the amplitude attenuation of light as it passes through the medium, accounting for the material’s absorption. An uncertainty when dealing with wind models of carbon stars is the chemical structure of amC and the corresponding optical properties. There are different ways  to produce and analyse amC dust in laboratories. A discussion of various methods is given in \citet{andersen1999}, for instance.  In the current paper, we are comparing the optical data set by \citet{rouleau1991} with that by \citet{jager1998} (sample cel 1000), denoted by \textit{RM}  and \textit{J10}, respectively. \textit{RM}  and \textit{J10}  represent two very different cases of commonly used data for amC dust in terms of extinction efficiencies (see bottom panel in Fig. \ref{fig:nk}). Comparing these two cases will allow us to judge the inherent uncertainties in the wind properties arising from differences in the optical data. Furthermore, \textit{RM}  has been used in previous DARWIN grids of C-stars, enabling comparisons to earlier results. A short summary of how the optical constants were derived for each chosen data set is given below.

\citet{rouleau1991} used the extinction data of submicron amC particles produced by \citet{bussoletti1987} to derive synthetic optical constants. The particles were obtained by striking an arc between two amC electrodes in a controlled Ar atmosphere and the resulting smoke was collected on quartz substrates. Spectroscopic measurements were performed directly on the substrates in the UV/visible and by scrapping the dust from the substrate and embedding it in KBr pellets in the IR. The total wavelength coverage extends from 1000\,Å to $300\,\mathrm{\mu m}$. From these measurements, \citet{rouleau1991} computed synthetic optical constants, which satisfied the Kramers-Kronig relations. 

\citet{jager1998} produced carbon materials by pyrolyzing cellulose at different temperatures ($400^{\circ}$C, $600^{\circ}$C, $800^{\circ}$C, and  $1000^{\circ}$C). The $\mathrm{sp^2/sp^3}$ ratio increases with temperature; thus, at $400^{\circ}$C, the pyrolyzed amC is more diamond-like and at $1000^{\circ}$C it is more graphite-like. The carbon samples were embedded in an epoxy resin and polished before the reflectance was measured in the range 200\,nm to $500\,\mathrm{\mu m}$. The complex refractive index, m, was derived from the reflectance spectra by the Lorentz oscillator method. 

The optical data, $n$ and $k$, of each sample are shown in Fig. \ref{fig:nk}. There is a clear difference in behaviour between the two samples. \textit{J10}  has a stronger wavelength dependence, while \textit{RM}  is almost constant over the wavelength range. The imaginary part, $k$, is higher for \textit{J10}  at all wavelengths compared to \textit{RM}. The $n-$data are lower for \textit{J10}  at wavelengths shorter than $2\,\mathrm{\mu m}$ and higher at wavelengths longer than $2\,\mathrm{\mu m}$.

\begin{figure}
\centering
      \includegraphics[clip,width=\hsize]{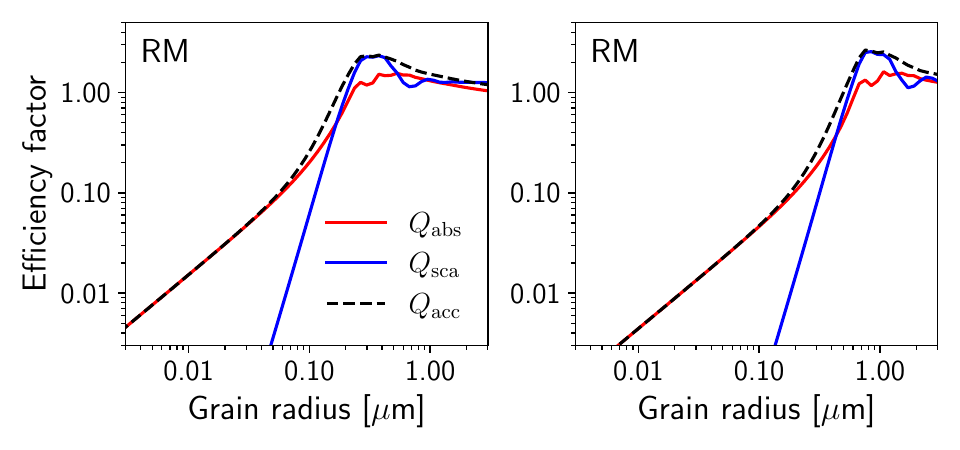}
      \\
      \includegraphics[clip,width=\hsize]{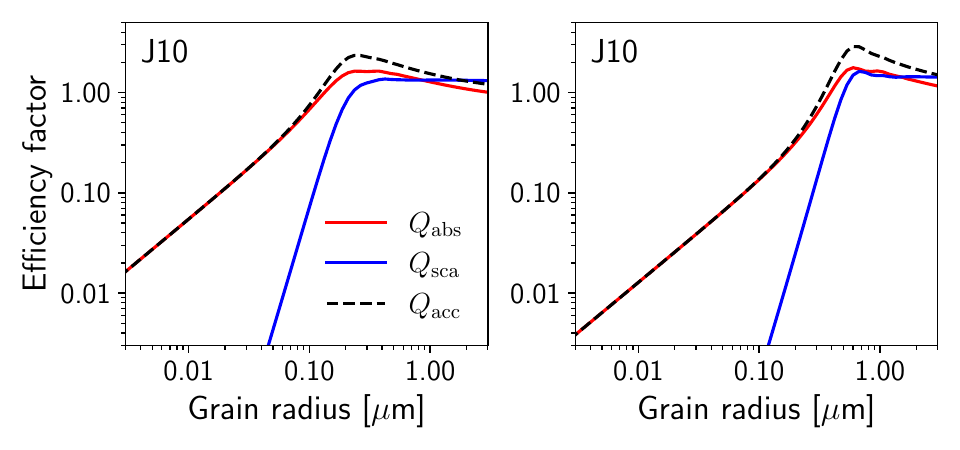}
  \caption{Efficiency factors as a function of grain size for an amC grain at different wavelengths; left panels: Values at $\sim1.2$ $\mathrm{\mu m}$ ($J$ band), right panels: Values at $\sim3.6$ $\mathrm{\mu m}$ (\emph{Spitzer} $[3.6]$ band). Red line: Absorption. Blue line: scattering. Black dashed line: Radiative pressure. The top and bottom panels represent the \textit{RM}  and \textit{J10}  optical data sets, respectively.}
  \label{fig:effgr}
\end{figure}

Figure \ref{fig:effgr} shows examples of efficiency factors (calculated from the optical constants) and their dependence on grain size for selected wavelengths (top panels: \textit{RM}, bottom panels: \textit{J10}). Absorption is the main contributor to radiation pressure for small grains, whereas scattering plays a role for grains with sizes that are comparable to the wavelength. The \textit{J10}  grains have higher absorption efficiencies than the \textit{RM}  grains, but the latter experience a more significant scattering boost and the resulting radiation pressure is comparable at larger grain sizes. 
%----------------------------------------------------------------- 
%                                                One column figure
%----------------------------------------------------------------- 
%-----------------------------------------------------------------
%                                                One column figure
%----------------------------------------------------------------- 
%-----------------------------------------------------------------
%--------------------------------------------------- One column table
\begin{comment}
   \begin{table}
    \caption[]{Optical data sets and their designations.}
    \label{tab:opac}
     $$
         \begin{array}{p{0.5\linewidth}l}
            \hline
            \noalign{\smallskip}
            Reference  &  Designation \\
            \noalign{\smallskip}
            \hline
            \noalign{\smallskip}
            \citet{rouleau1991} & \textit{RM}  \\
            \citet{jager1998} cel1000 & \textit{J10}  \\
            \noalign{\smallskip}
            \hline
         \end{array}
     $$
   \end{table}
\end{comment}

\subsection{Grid parameters}
\label{sec:gridparam}

To evaluate how the dynamic structures and outflows of C-type AGB stars are affected by different dust optical properties, we calculated two grids of DARWIN models with either \textit{RM}  or \textit{J10}  optical data. The combinations of stellar parameters that were modelled are based on the grid presented in \citet{eriksson2023} with extended temperature and luminosity ranges. The carbon excess (i.e. the carbon that is not bound in CO molecules) represents the available raw material for the wind-driving dust species and is calculated by $\mathrm{log(\varepsilon_C-\varepsilon_O)+12}$, where $\mathrm{\varepsilon}$ is the abundance by number of the element.
Our models include carbon excesses of 8.2, 8.5, and 8.8, which correspond to a C/O ($\mathrm{\varepsilon_C/\varepsilon_O}$) ratio of 1.35, 1.69, and 2.38, respectively, with the adopted chemical composition by \citealt{asplund2005} (except for carbon). For a discussion on the effects of different overall metallicity on DARWIN models, we refer to \cite{bladh2019a}.

Stellar pulsation is simulated by a sinusoidal variation of the position of the innermost mass shell (see Sect. \ref{sec:methoddarwin}), with a period that is set according to the period-luminosity relation presented in \citet{feast1989}, and the luminosity varies in phase with the expansion and contraction of the star.
The velocity amplitudes at the inner boundary of $\mathrm{\Delta u_p=2}$, 4 and 6\,$\mathrm{km\,s^{-1}}$ result in shock amplitudes of about $15-20\,\mathrm{km\,s^{-1}}$ in the inner atmosphere, which is in good agreement with observations (see \citealt{nowotny2010a} and references therein).  Table \ref{tab:grid} gives a complete list of input parameter combinations.

\subsection{Synthetic spectra and photometry}
\label{sec:synth}

The time series of snapshots extracted from the DARWIN models provide information about the atmosphere and wind properties as a function of radial distance from the star. For the detailed a posteriori radiative transfer calculations, snapshots covering three to four periods at two different epochs of the simulation were used. The calculations were performed on a selection of models in the grid (details on selection in Sect \ref{subs:modspectra}) using the \texttt{COMA} code \citep{aringer2016,aringer2019}. Based on the resulting synthetic opacity sampling spectra ($\mathrm{R=10\,000}$), we computed the photometric filter magnitudes following the Johnson-Cousins \emph{BVRI} system \citep{bessell1990} and the Johnson-Glass \emph{JHKLL’M} system \citep{bessell1988}. We refer to \citet{nowotny2011} (in particular Sect. 2.3) for more details on the procedure. The transmission functions and zero-points for \textit{Spitzer, Gaia,} and \textit{2MASS} filters were retrieved from the SVO Filter Profile Service\footnote{\url{http://svo2.cab.inta-csic.es/theory/fps/index.php}}.

%--------------------------------------------------------------------
\section{Results: Effects of the optical data}
\label{sec:resultmodel}
In this section, the resulting differences between the two model grids are presented. We first discuss the time-averaged dynamical properties of both grids, followed by the photometric properties of a selection of the models. An overview of general trends as well as systematic deviations is given. 

%-------------------------------------------------------------
%                                             Simple A&A Table
%-------------------------------------------------------------
%
\begin{table}
\caption{Input parameter combinations covered by the two model grids, where $\Delta$ denotes the grid step.}             % title of Table
\label{tab:grid}      % is used to refer this table in the text
\centering                          % used for centering table
\begin{tabular}{c c c c c}        % centered columns (4 columns)
\hline\hline                 % inserts double horizontal lines
\noalign{\smallskip}
$\mathrm{M_{\star}}$ & $\mathrm{T_{\star}}$ & $\mathrm{logL_{\star}/L_{\odot}}$ & $\mathrm{\Delta u_p}$ & $\mathrm{log(C-O)}$ \\    % table heading 
$\mathrm{[M_{\odot}]}$ & $\mathrm{[K]}$ & & $\mathrm{[km\,s^{-1}]}$ & $+12$ \\
\noalign{\smallskip}
\hline                        % inserts single horizontal line
\noalign{\smallskip}
    & $\Delta=200$ & $\Delta=0.15$ & $\Delta=2$ & $\Delta=0.3$ \\ \noalign{\smallskip}
   0.75 & $2400-3200$ & $3.55-3.85$ & $2-6$ & $8.2-8.8$ \\
   1.0  & $2400-3200$ & $3.70-4.00$ & $2-6$  & $8.2-8.8$\\
   1.5  & $2400-3200$ & $3.85-4.15$ & $2-6$  & $8.2-8.8$\\
   2.0  & $2400-3200$ & $3.85-4.15$ & $2-6$  & $8.2-8.8$\\
\hline                                   %inserts single line
\end{tabular}
\end{table}

%                                                One column figure
%----------------------------------------------------------------- 
\begin{figure}
   \centering
   \includegraphics[width=\hsize]{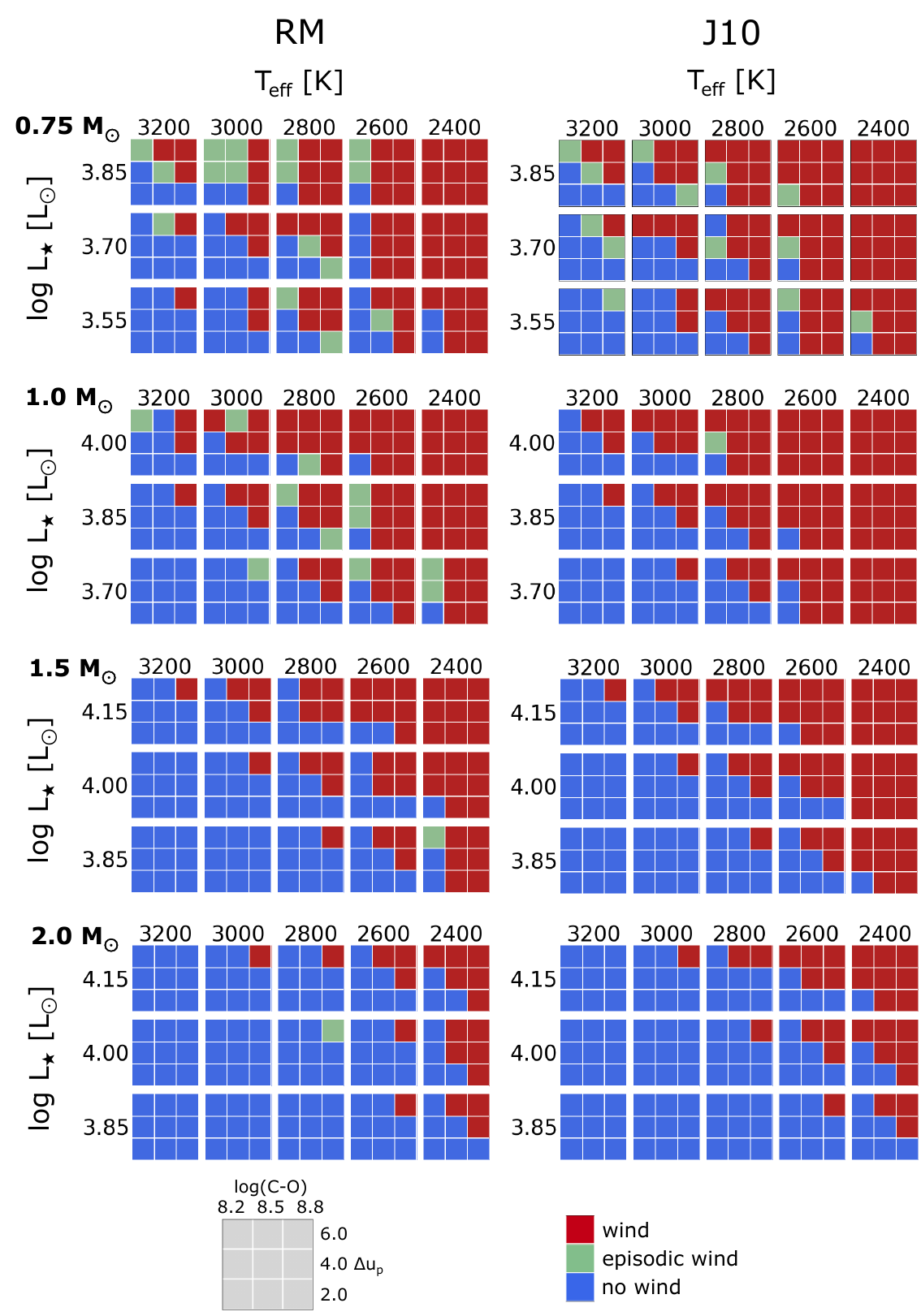}
      \caption{Schematic overview of the dynamic behaviour of the models as a function of input parameters. \emph{Left column:} Wind maps obtained with \textit{RM}  opacity data. \emph{Right column:} Wind maps with \textit{J10}  opacity data. \emph{From top to bottom:} Wind maps for 0.75, 1.0, 1.5, and 2.0 solar masses. The colours represent dynamic behaviour, and each temperature and luminosity combination is further divided into squares indicating piston velocity, $\Delta u_p$, and carbon excess, log($\mathrm{C-O}$). See bottom legend for details.}
         \label{fig:windmaps}
\end{figure}

%-----------------------------------------------------------------

\subsection{Dynamical properties}
\label{subs:dyn}

In Fig. \ref{fig:windmaps}, an overview of the dynamical behaviour of each model is shown as a function of input parameters. These so-called wind maps (first used in \citealt{eriksson2014} for analysing DARWIN grids) are arranged in a format resembling a Hertzsprung-Russel diagram (HRD) with temperatures increasing to the left. Each square represents a unique model and is coloured by the corresponding dynamical properties. Here, red squares represent models with a stellar wind, blue squares represent models with no wind, and green squares represent models with episodic mass-loss. The panels from top to bottom illustrate wind properties for models with stellar masses of 0.75, 1.0, 1.5, and 2.0 $M_\odot$, while the left and right columns represent the results from the \textit{RM}  and \textit{J10}  optical data sets, respectively. 
A total of 540 models were computed for each data set, out of which 206 of the \textit{RM}  models resulted in winds, 29 showed episodic behaviour and 305 models did not produce outflows. The corresponding numbers for \textit{J10}  are 240 with winds, 14 with episodic behaviour and 286 with no outflow. The \textit{J10}  dust opacities are, hence, more favourable for sustaining an outflow. As a result, they affect the location of the boundary in the wind maps between models with and without wind, based on stellar parameters. Furthermore, the wind maps 
clearly illustrate how each input parameter affects the winds, namely, that mass loss is generally favoured by low $T_{\star}$ (easier to form dust at cooler temperatures), high $L_{\star}$ (more momentum transferred from radiation to dust grains), small $M_{\star}$ (shallower potential well), large $\mathrm{C-O}$ (more free carbon to form amC grains), and a large $\Delta u_p$ (stellar layers reach out to a greater distance from the centre of the star during pulsations),
as previously concluded in, for instance, \citealt{mattsson2010,bladh2019a}.
The results from the dynamical modelling are compiled in Appendix \href{https://doi.org/10.5281/zenodo.14904461}{A}, including a comparison of the mean properties of the dynamical models shown in a wind map format. The colour of each square is given by the \textit{J10}/\textit{RM}  ratio, where blue colours indicate \textit{RM}$>$\textit{J10} and red colours \textit{J10}$>$\textit{RM}. The wind maps show that models that are diverging from the general trends are mainly situated at the boundary between the wind and no-wind regions of the parameter space. 

\begin{figure}
   \centering
   \includegraphics[width=\hsize]{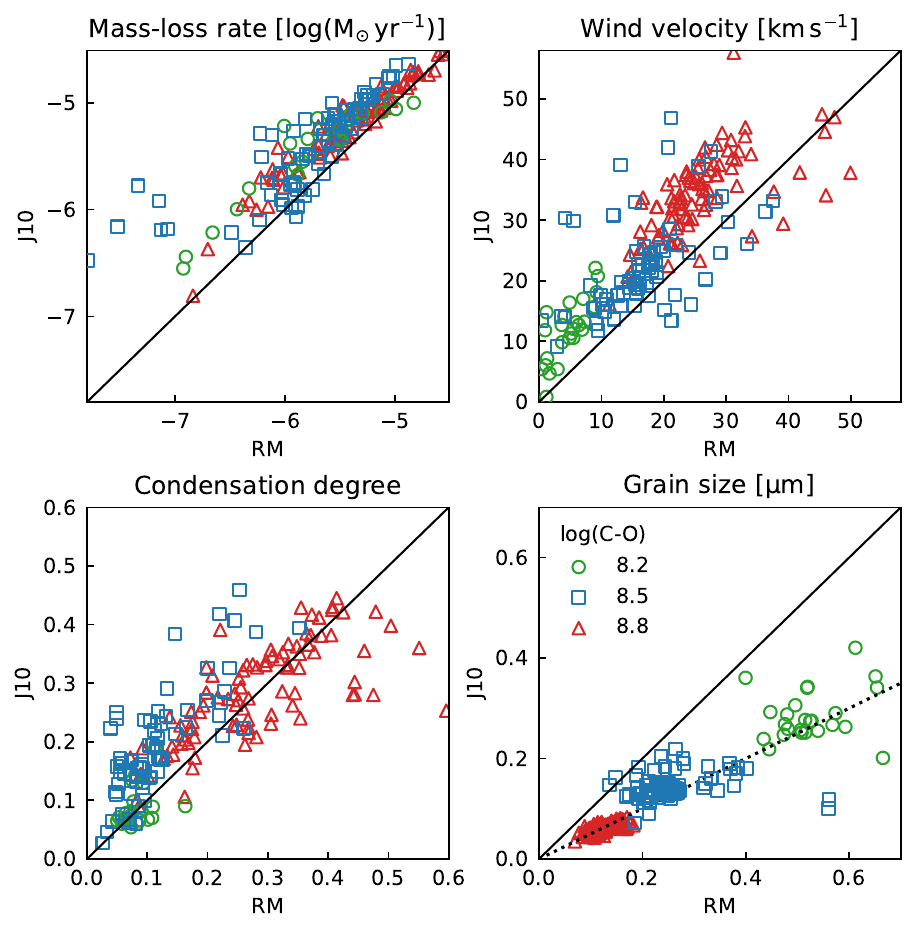}
      \caption{Comparison of wind and grain properties predicted by
        the models. The x-axis represent the results from models based
        on \textit{RM}  opacity data and the y-axis represent the results based
        on \textit{J10}  opacity data. The colours indicate the amount of excess
        carbon (see legend in the bottom right panel). Only models with
        steady winds are depicted.}
         \label{fig:dustwindprop}
\end{figure}

Figure \ref{fig:dustwindprop} shows a comparison of the dynamic properties between the two grids for models where an outflow was reached and a mean could be calculated. 
The dynamical properties are time averages for models where the outflow reaches the outer boundary of 25 stellar radii (definition of a wind model, see Sect. \ref{sec:methoddarwin}). For most models, a few hundred periods were used to determine the means. For some combinations of stellar parameters, the outflow intermittently reaches the outer boundary, resulting in episodic mass loss. Following \citet{eriksson2023}, a model is defined as episodic when wind conditions prevail for more than 15\% but less than 85\% of the total time interval (after the first instance of wind conditions).
 The plots in Fig. \ref{fig:dustwindprop} are colour-coded according to carbon excess as shown in the bottom right panel. Looking at the wind properties (upper two panels), there is a trend toward higher wind velocities and also to somewhat higher mass-loss rates for the \textit{J10}  grid. 
Figure \href{https://doi.org/10.5281/zenodo.14904461}{A.2} shows that the mass-loss rates of most \textit{J10}  models are higher than those of the \textit{RM}  models, typically within a factor of 2. The outliers (with values less than $10^{-7}\,M_\odot\,yr^{-1}$ for \textit{RM}) are models located next to the no-wind boundary, where dust opacities become more decisive in terms of driving the winds effectively.
 The wind velocities show an expected trend related to the carbon excess where more available carbon leads to higher velocities (more dust gives more acceleration to the outflow). 
The grain properties are displayed in the bottom two panels. The condensation degree shows no systematic dependence on the chosen optical data, except for models with carbon excess falling within the middle range; namely, $\mathrm{log(C-O)=8.5}$, where the \textit{J10}  models exhibit a higher condensation degree compared to the \textit{RM}  models. In general, the grain sizes become smaller with increasing carbon excess. The \textit{J10}   grains are on average about half the size of the \textit{RM}  grains as illustrated by the dotted line in the figure.

\begin{figure}
    \centering
    \includegraphics[width=\hsize]{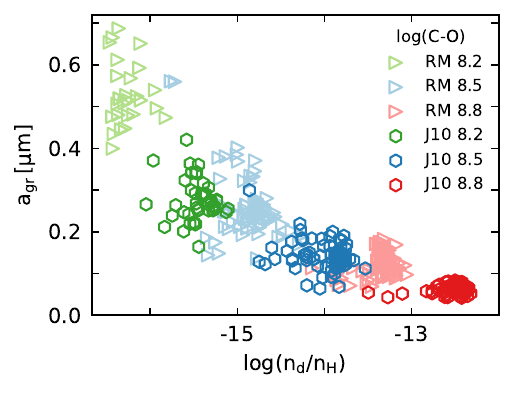}
    \caption{Average grain size versus the abundance of grains by number relative to hydrogen. The anti-correlation of these two quantities is a consequence of the competition between nucleation and grain growth (see text for details).}
    \label{fig:ndnh_agr}
\end{figure}

The sizes of the wind-driving amC grains result from an intricate interplay of different processes. During the early phases of condensation, when the grains are small, there is a competition between nucleation (i.e. the formation of new dust particles from the gas) and the growth of existing grains. Eventually, as the collective surface of the existing grains grows, the depletion of carbon in the gas phase is dominated by the latter process and nucleation becomes ineffective. For the \textit{J10}  models, this happens a bit later, as the steeper dependence of the absorption coefficient on wavelength (see Fig. \ref{fig:nk}) leads to higher grain temperatures, delaying efficient grain growth compared to the \textit{RM}  case. Therefore, more seed particles are formed in the \textit{J10}  models, eventually resulting in similar (or even higher) values of the condensation degree, despite the smaller final sizes of the grains.
The anti-correlation between the grain sizes and the abundances of grains, resulting from the competition between nucleation and grain growth, is illustrated in Fig. \ref{fig:ndnh_agr}.

Since the \textit{J10}  dust grains are more opaque (higher mass absorption coefficient), as seen in, for instance,  Figs.\ref{fig:nk} and \ref{fig:effgr}, they will absorb the stellar radiation more efficiently and accelerate the outflow at smaller grain sizes compared to the \textit{RM}  dust. As the dust-gas mixture moves outwards, the grain growth will slow down due to rapidly falling densities. 
However, there are several models in the carbon excess middle range where grain sizes are more similar between the two data sets. This can be explained by a scattering boost that occurs for grains around $0.2-0.3$ microns (which are the approximate sizes of the relevant \textit{RM}  grains). For grains in this size range, scattering gives an extra boost to the radiation pressure (see Fig. \ref{fig:effgr}). The grains are consequently accelerated faster away from the star to distances where the grain growth may slow down or stop completely due to the changed conditions (lower densities) in the surrounding environment.

\subsection{Photometric properties} \label{subs:modspectra}
A posteriori detailed radiative transfer calculations were performed on a selection of models from the grid. To ensure a representative selection,  models were chosen on the basis of
\begin{enumerate}[i]
  \item covering a variety of stellar parameters;
  \item the results of the dynamical modelling, including, for instance, models that follow the trends as well as some outliers and models where the qualitative dynamics (i.e. wind, episodic, or no wind) differ between the \textit{RM}  and \textit{J10}  grid, also including models without wind for photometric comparisons;
  \item covering a broad range in colours (based on the photometric result from previous papers, \citealt{eriksson2014,eriksson2023}).
\end{enumerate}

%----------------------------------------------------------------- 

\begin{figure}
    \centering
    \includegraphics[width=\hsize]{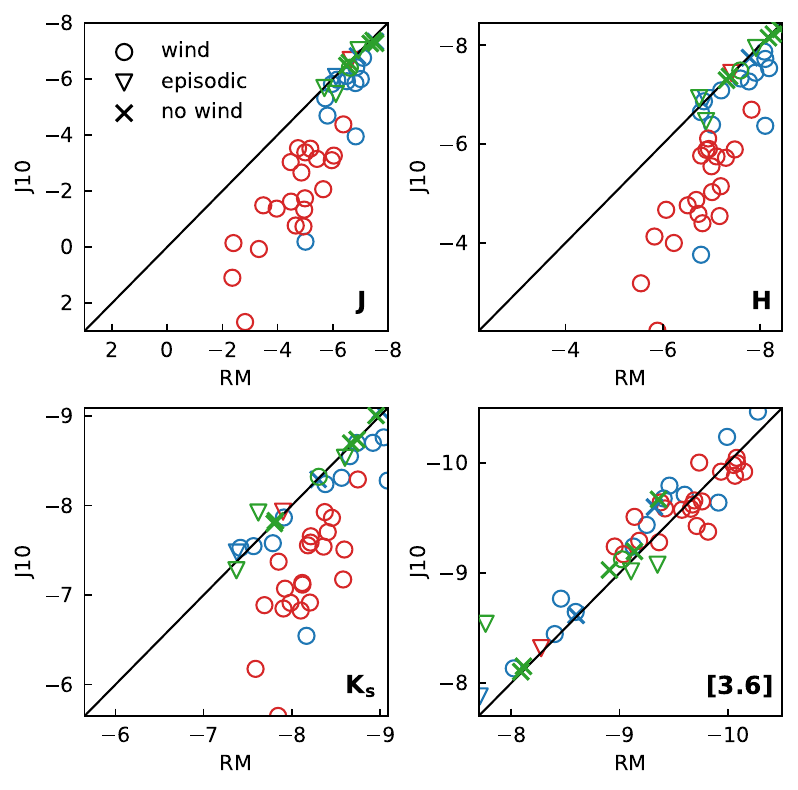}
    \caption{Mean $J$, $H$, $K_s$, and $\mathrm{[3.6]}$ magnitudes for the model selection (\textit{J10} vs \textit{RM}). Symbols are explained in the top left panel, colours of symbols indicate carbon excess, same as in Fig. \ref{fig:dustwindprop}.}
    \label{fig:vjhk}
\end{figure}

%----------------------------------------------------------------- 

\noindent The final selection consists of 46 models, listed in Table \href{https://doi.org/10.5281/zenodo.14904461}{B.1} giving the mean $J$, $H$, $K_s$, and $\mathrm{[3.6]}$ magnitudes for the \textit{RM}  and \textit{J10}  data sets. In Appendix \href{https://doi.org/10.5281/zenodo.14904461}{B}, the dynamical properties are plotted for the selected models for clarity and easier reference. The photometric properties are compared in Fig. \ref{fig:vjhk}. The colours of the symbols indicate carbon excess and the shapes mark dynamical properties as shown in the top left panel. We note the differing magnitude scales of the panels. In general, we note that both versions of the models tend to get fainter in the $J$, $H,$ and $K_s$ filters with increasing carbon excess; namely, with more dust contributing to circumstellar extinction and reddening. The flux at the stellar photosphere is independent of the choice of dust optical data and the low-C/O models with little or no dust in the circumstellar environment line up close to the one-to-one line in all panels.
In contrast, models with high carbon excess (but also some in the middle range) show a significant difference for the two dust data sets in the $J$, $H,$ and $K_s$ magnitudes. For these models, the differences range between two magnitudes in the $J-$ and $H-$band and about one magnitude or more in the $K_s-$band. The divergent models correlate with higher degrees of condensation (for both \textit{RM}  and \textit{J10}). In other words, these models are enshrouded in dust, and with increasing amounts of dust in the CSE, the effect of dust opacities becomes larger. 
The $[3.6]$ filter (bottom-right panel) shows a qualitatively different behaviour, with all models aligning closely to the one-to-one line, regardless of the amount of carbon excess or the presence of a wind. The $[3.6]$ band happens to be in the spectral region where the dominant effect of dust changes from absorbing stellar flux to emitting thermal radiation.

%-----------------------------------------------------------------
\begin{figure}
    \centering
    \includegraphics[width=0.9\hsize]{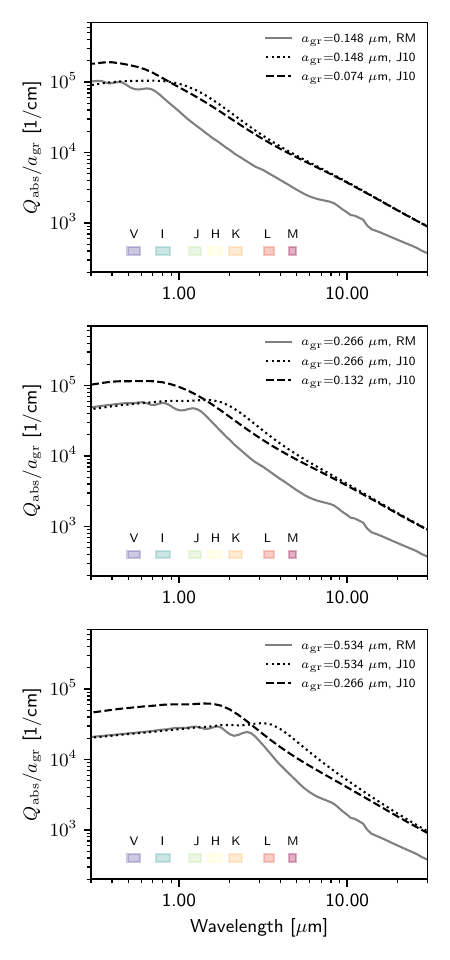}
    \caption{$Q_{abs}/a_{gr}$ as a function of wavelength for \textit{RM}  (solid line) and \textit{J10}  (dashed and dotted) optical data sets and several grain sizes as indicated in the legend. See text for details.}
    \label{fig:qda_mix}
\end{figure}
%-----------------------------------------------------------------

To understand the reasons for the differences in the $J$, $H,$ and $K_s$ filter magnitudes between models with \textit{RM}  and \textit{J10}  optical data, we consider the corresponding mass absorption coefficients. From Eq.(\ref{eq:opacityRe}), we see that they depend on the product of $Q'_{\rm abs} = Q_{\rm abs}(a_{\rm gr}, \lambda)/a_{\rm gr}$ (efficiency factor divided by grain radius) and the total amount of condensed material in all grains at a given location ($f_{c} \propto K_3/\rho$). For most models, the degree of condensation is equal or higher in the \textit{J10}  case (Fig. \href{https://doi.org/10.5281/zenodo.14904461}{B.1}). Figure \ref{fig:qda_mix} illustrates the dependence of $Q'_{\rm abs}$ on wavelength for both materials and several grain sizes. Comparing the \textit{RM}  and \textit{J10}  data for the same grain size (full and dotted lines), we see that the \textit{J10}  values are higher than \textit{RM} -- except for the shortest wavelengths, where they are about equal (large particle limit, efficiency factors approaching unity). This is consistent with the results shown in Figs. \ref{fig:nk} and \ref{fig:effgr}. However, as explained above (Sect. \ref{subs:dyn}, Fig. \ref{fig:dustwindprop}),  the grains tend to have smaller radii in the \textit{J10}  models, typically half of the values of the \textit{RM}  models. Applying a factor of 0.5 to the grain radius increases $Q'_{\rm abs}$ and, in that case, the \textit{J10}  values are higher than \textit{RM}  at all wavelengths (dashed versus full lines). We also note that the two curves for \textit{J10}  approach each other at long wavelengths, as expected, since $Q'_{\rm abs}$  becomes independent of the grain radius in the small particle limit. In summary, both the optical properties of the individual grains and the total amount of dust contribute to higher absorption in the \textit{J10}  models, explaining the trend towards fainter near-IR magnitudes.

For spherical dust grains, a factor of 2 increase in the grain radius means a factor of 8 increase in volume. Thus, if two corresponding models show similar condensation degrees, but differ in grain size by a factor of 2, the model with smaller grains contains more dust particles. A dust-enshrouded model with more abundant, smaller, and more opaque grains will absorb more stellar light, which leads to lower near-IR (NIR) fluxes.

%----------------------------------------------------------------- 

\section{Discussion}
\label{sec:resultobs}

In this section, we compare our results with observational data and other models from the literature. We also discuss the implications of specific model assumptions.

\subsection{Comparisons to observational data}
\label{sec:compobs}

Before we start our comparison with various types of observational data, it is important to point out that the modelling results discussed here represent grids (or sub-samples of grids, for photometry). This means that equal weight is given to all combinations of stellar and pulsation parameters, in contrast to stars on evolution tracks or a distribution of parameters corresponding to a stellar population. Therefore, we do not expect to match the observations in a statistical sense, but we check if the models cover the observed ranges and if trends with model properties are in accordance with observations. In the present context, a key question is if there are significant differences for the \textit{RM}  and \textit{J10}  grids that indicate a preference for using one of the optical data sets.

\begin{figure}
    \centering
    \includegraphics[width=\hsize]{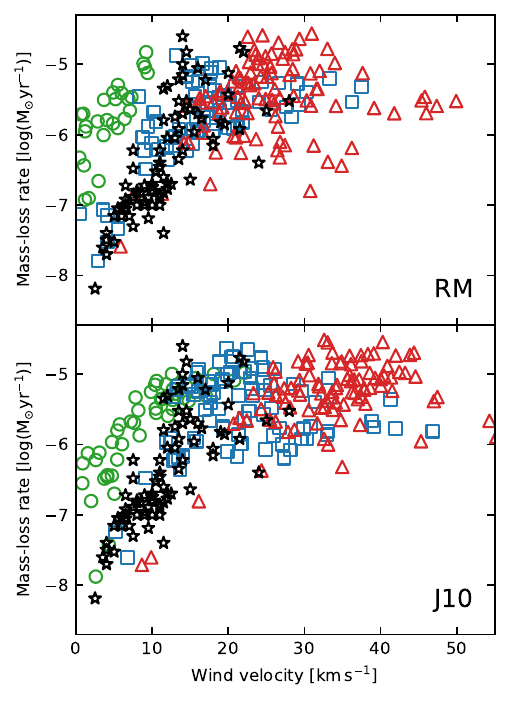}
    \caption{Mass-loss rates vs wind velocities for DARWIN models at different carbon excess (colours and symbols explained in Fig. \ref{fig:dustwindprop}), with the corresponding observed properties shown as black stars (data from \citealt{schoier2001,ramstedt2014,danilovich2015}). }
    \label{fig:windmodobs}
\end{figure}

The mass-loss rate and wind velocity of AGB outflows can be estimated from mm-wave observations of CO rotational lines and associated radiative transfer modelling. In Fig. \ref {fig:windmodobs}, we compare the observed properties of carbon stars with the corresponding values from our grids (top panel: \textit{RM}  grid, bottom panel: \textit{J10}  grid). The observations were retrieved from \citet{schoier2001} and \citet{ramstedt2014}, the latter including updated values for sources in common, and from \citet{danilovich2015}. The noted trend from the previous section is clearly visible, namely, the outflows from the \textit{J10}  models typically reach higher velocities. This is, for example, illustrated by models with the lowest carbon excess (green circles) where the \textit{RM}  models reach a maximum velocity of about $\mathrm{10\,km\,s^{-1}}$ while the corresponding \textit{J10}  models reach over $\mathrm{20\,km\,s^{-1}}$. 
As a result, this group of models for the \textit{J10}  grid overlaps with the observed sources, particularly those characterised by high mass-loss rates, while the \textit{RM}  grid lies outside the observed range. For both grids, however, models with a moderate carbon excess (blue squares) are in best agreement with the range of the empirical data.

%----------------------------------------------------------------- 
\begin{figure*}
    \centering
    \includegraphics[width=\textwidth]{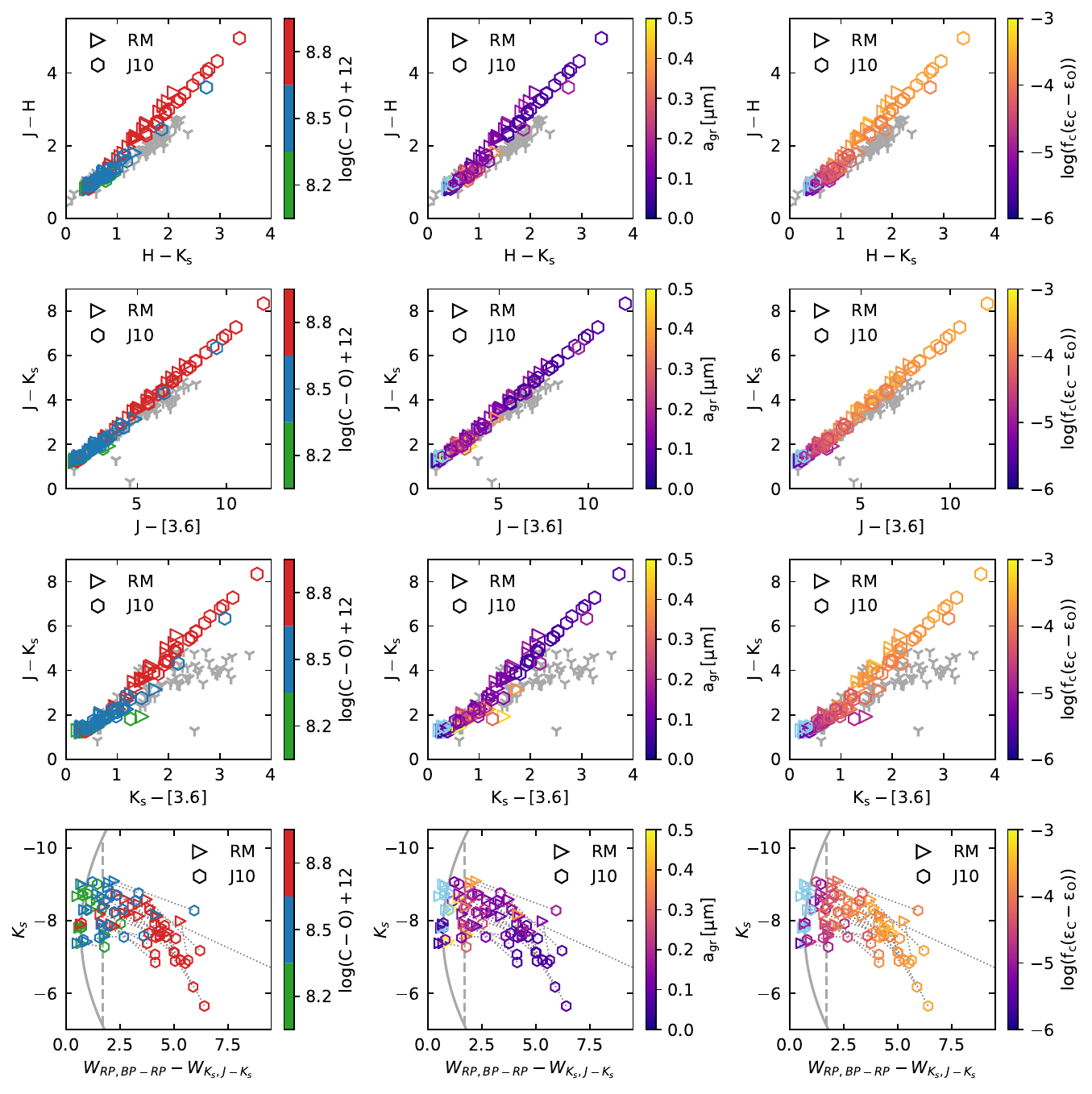}
    \caption{Colour-colour-diagrams (CCDs) given in rows 1-3, as well as a \emph{Gaia-2MASS} diagram (see text for details). All three columns show the same model data, but colour-coded by different properties: carbon excess (input parameter; left column), resulting grain size (middle), and amount of dust (condensation degree times $\mathrm{C-O}$ abundance; right). Model symbols in light blue indicate models with no wind. Observations are plotted as grey symbols in the CCDs (top three rows), drawn from \citet{jones2017}. }
    \label{fig:ccds}
\end{figure*}
%-----------------------------------------------------------------

Another useful way to compare models with observations is to construct colour-colour diagrams (CCDs). We note that for these comparisons, we used the results of a posteriori radiative transfer that was only applied to a subsample of the models (explained in Sect. \ref{subs:modspectra}). 
In Fig. \ref{fig:ccds}, we plot three CCDs ($J-H$ vs $H-K_s$, $J-K_s$ vs $J-[3.6],$ and $J-K_s$ vs $K_s-[3.6]$ in rows 1-3) as well as a \emph{Gaia-2MASS} diagram (explained below). 
The three columns show the same data, but the models are colour-coded differently, according to carbon excess, grain size, and dust abundance (i.e. the degree of condensation multiplied with the abundance of C not bound in CO). The latter quantity (see right column) shows a clear correlation with the NIR colours in all diagrams, which is expected, as the amount of dust in the wind largely defines circumstellar reddening. In contrast, there is no obvious trend with grain size (middle column), apart from the fact that the reddest models (mostly from the \textit{J10}  grid) tend to have the smallest grains. This can be explained by their high carbon excess (see left column), leading to efficient nucleation, namely, a large abundance of grains (see Sect. \ref{subs:dyn}).

Observations are plotted in the three CCDs for comparison and
 are retrieved from the C-AGB sample of \citet{jones2017}, which includes 148 carbon-rich AGB stars in the Large Magellanic Cloud (LMC). We note, as discussed in \citealt{bladh2019a}, different metallicities do not significantly affect the mass-loss or, consequently, the circumstellar reddening by dust -- allowing for the comparison of models based on solar values with observations in the LMC. The sample of stars ranges from relatively blue to ‘very’ and ‘extremely’ red objects (based on the overall shape of the spectra). 
 When comparing the distributions of the observed values with those of the synthetic ones, we find a good general agreement, but we note that some of the reddest models fall outside the observed ranges. As such, this is not a reason for concern, as we are dealing with grids that span a wide range of stellar and pulsation parameters, in contrast to synthetic stellar populations, constrained by evolution models. 
 Given the large parameter space explored (and assuming that the selection of models described in Sect. \ref{subs:modspectra} and Appendix \href{https://doi.org/10.5281/zenodo.14904461}{B}) is representative of the whole grid, regarding NIR properties), we could expect that the synthetic colours cover the whole range of observed values. However, this is not the case, which is intriguing, as it may point to shortcomings in the assumptions about dust formation in the underlying models. In this context,  we discuss below the effects of using a classical nucleation rate, in contrast to assuming a given seed particle abundance, as done in other models.

A defining feature of the DARWIN models is a self-regulating feedback between the dynamics, grain growth, and the optical properties of the grains, such that they predict combinations of mass-loss rates, wind velocities, and grain sizes, for given stellar parameters and micro-physical data. This is in contrast to studies using grids of synthetic spectra based on parameterised models (see e.g. \citealt{nanni2019b, groenewegen2018} and references therein), where input quantities such as stellar parameters, wind properties, dust grain sizes, and dust optical properties can be varied independently, to allow a fitting to observational data. 
In the DARWIN models presented in Sect. \ref{sec:resultmodel}, the formation of new dust grains from the gas is treated with a classical nucleation rate, determining the abundance of grains in competition with the growth of existing grains (see Sects.~\ref{sec:methoddarwin} and \ref{subs:dyn}). However, it is unclear if classical nucleation theory is a realistic assumption in the context of carbon star atmospheres. Therefore, we have produced several grids of test models where the grain abundance is instead set to a fixed value, as described in 
Appendix ~\href{https://doi.org/10.5281/zenodo.14904461}{C}. 
The chosen values (grain abundance by number, relative to hydrogen: $10^{-15}$, $10^{-13}$ and $10^{-11}$) range from typical values found in the DARWIN models, using a classical nucleation rate, to a regime where resulting grain sizes are within the small particle limit regarding NIR optical properties (see Figs.~\ref{fig:ndnh_agr} and \href{https://doi.org/10.5281/zenodo.14904461}{C.1}).
With increasing seed particle abundance, the resulting grain sizes decrease, as expected, and condensation degrees tend to be higher, except for models with $n_d/n_{\rm H} = 10^{-15}$. Wind velocities also tend to increase with higher seed particle abundance. Most relevant for the current study is that the mass loss rates are not strongly affected, although there are some differences, depending on the carbon excess (see Figs. \href{https://doi.org/10.5281/zenodo.14904461}{C.2-C.4}).
We find that for $n_d/n_{\rm H} = 10^{-15}$ (giving grains close to the typical sizes in the models with nucleation rates), the test models best match both the empirical dynamical properties (mass loss rate vs wind velocity) and the observed colours, although they do not stretch all the way to the objects with most circumstellar reddening. For the other two cases, with smaller grains, the models partly show significant deviations from the empirical values.

These results are in contrast to the findings of \citet{nanni2016} and \citet{nanni2019}, who obtained the best match for photometry of carbon stars in the SMC and LMC with high values of seed particle abundances, resulting in small grains. In this context we note, that those models differ in several decisive ways from the models presented here. First, they assume steady-state winds (time-independent outflows, neglecting all effects of stellar pulsation), while using simpler radiative transfer to determine temperatures and radiative acceleration. Furthermore, there are a number of differences in micro-physical assumptions that affect grain growth. Finally, the steady-state wind models are based on stellar parameter combinations from stellar evolution tracks (rather than a grid giving equal weight to all models, as in the present work), but with the mass-loss rate as an input parameter (not a result) of the wind modelling. Therefore, the physical constraints defining radial structures and dust grain properties are very different, compared to DARWIN models. It should be mentioned that \citet{nanni2016} also used CCDs involving longer wavelengths than those shown in Fig. \ref{fig:ccds}. However, we are currently not considering photometry beyond the 3.6 band, due to known problems with the C$_3$ opacity data (see \citealt{aringer2019}) and the fact that the models do not include SiC dust at present. These disparities would thus make a comparison with observations in the MIR problematic. 

\citet{lebzelter2018} designed a diagram to separate O-rich and C-rich AGB stars (here referred to as the \emph{Gaia-2MASS} diagram, see bottom row of Fig. \ref{fig:ccds}). They defined the Wesenheit function $W_{RP}$ (a reddening-free combination of \emph{Gaia} $G_{BP}$ and $G_{RP}$ magnitudes) and combined it with the near-infrared Wesenheit function $W_{K_s,J-K_s}$  (used to obtain a reddening-free magnitude for red giants) to create a 2D diagram by plotting it against the $K_s$ magnitude. 
The curved line separates O-rich and C-rich AGB stars (C-rich are located to the right of this line).
All models in our subsample fall in the carbon-star region of the diagram except a few no-wind and episodic models with \textit{RM}  dust opacities. When changing to \textit{J10}  opacities, the model generally moves to the right in the figure (often into the region labelled 'extreme C-rich stars' in \citealt{lebzelter2018}, i.e. the region to the right of the dashed line). To illustrate this, corresponding \textit{RM}  and \textit{J10}  models are linked by dotted lines. In other words, a given observed source could be matched by models with smaller $\mathrm{C-O}$ values when using the \textit{J10}  dust opacities.

\subsection{Model assumptions: Motivations and consequences}
\label{sec:modelass}
The model results discussed in Sect. \ref{sec:resultmodel} are based on a number of assumptions that make dust formation and wind driving relatively efficient, compared to other studies in the literature. One of these assumptions concerns the so-called sticking coefficients (or reaction efficiency factors), namely the probability that a gas particle hitting a dust grain will stick to the surface, contributing to grain growth. In the  grids used in the present study, we chose to set these coefficients to 1 to keep the models comparable to earlier grids (e.g. \citealt{mattsson2010,eriksson2023}). Lowering the values would result in lower degrees of condensation, lower wind velocities, lower mass loss rates, and, consequently, less circumstellar reddening. The effects would be stronger in the stellar parameter regime close to the wind-driving border line. Some authors have argued for values of sticking coefficients that would be significantly below unity from a micro-physical point of view (see e.g. the discussion in \citealt{andersen2003}). Recent 3D model results, however, have suggested that 1D spherical models may have the tendency to underestimate the dust formation efficiency from a morphological point of view. In the 3D RHD ‘star-and-wind-in-a-box’ models from \citet{freytag2023}, the combined effects of pulsation and large-scale convective motions lead to the formation of clumpy gas clouds in the stellar atmosphere. Dust condensation occurs preferentially in these clouds, closer to the stellar surface than the spherical averages of temperatures would indicate. This can lead to a dust-driven outflow for stellar parameters, where a spherical model will not develop a wind. Choosing high values for the sticking coefficients in 1D spherical models can be seen as a way to compensate for this 3D effect. 

Another model assumption that may affect the mass loss is that the dust grains remain position-coupled to the gas up to the outer boundary of the models. The efficiency of dynamical coupling between the gas and dust (i.e. the drag force) depends on collision rates and, therefore, on the density. As density, on average, decreases with distance from the star, the grains will eventually decouple from the gas and be accelerated outwards more strongly, since they only represent a much smaller part of the total mass. It was shown by \citet{sandin2003}, that the effects of drift in a time-dependent wind model can be quite complex, since the grains spend much time in the dense regions behind shock waves, where drift velocities are low, before they quickly cross the low-density regions between these dense layers. A multi-fluid approach is computationally much more demanding than position-coupled models. So far, only a small number of time-dependent wind models for C-type AGB stars that include drift between dust and gas have been published (e.g. \citealt{sandin2020}). Therefore, it is difficult to estimate the overall effect on mass loss and observables at present. In the DARWIN models discussed in this paper, we have chosen to keep the assumption of position coupling between gas and dust for the sake of computational efficiency, making it possible to span wide ranges in stellar parameters.

Having discussed these caveats, we note that in this study we are mainly interested in the relative effects of different dust optical data sets available in the literature -- and not as much in absolute values of the resulting wind properties (as long as they fall within the observed ranges). Nevertheless, certain basic assumptions and their consequences for the model results will have to be explored further in the future.

\section{Summary and conclusions}
\label{sec:sum}
Amorphous carbon (amC) grains play a crucial role in driving the winds of C-rich AGB stars. We have investigated how the predicted wind properties of DARWIN models are influenced by the choice of optical properties of the amC dust. We  used the $n$ and $k$ data of \citet{rouleau1991} and \citet{jager1998} (cel 1000) in our comparison and computed two extensive grids of models. Each grid contains 540 combinations of fundamental stellar parameters (i.e. stellar mass, effective temperature, luminosity, carbon excess) and pulsation properties.

The \textit{J10}  grid resulted in more models with wind and fewer models with episodic behaviour than the \textit{RM}  grid. This suggests that the \textit{J10}  grains are more efficient at driving and sustaining a wind. On average, the \textit{J10}  models show a small but noticeable tendency towards higher mass-loss rates than the \textit{RM}  models (typically within a factor of two), with only a handful of exceptions. The \textit{J10}  dust grains are more opaque and more efficient absorbers of stellar radiation. They start accelerating outwards at smaller sizes compared to the \textit{RM}  particles and consequently leave the growth zone sooner. An approximate two-to-one relation between the \textit{RM}  and \textit{J10}  grain sizes was visually identified. While the \textit{J10}  grains tend to be smaller, there are more of them being formed, resulting in similar degrees of carbon condensation. In the competition between nucleation (i.e. the formation of new grains from the gas) and the growth of existing grains, the latter is less efficient due to dust temperatures being higher in the \textit{J10}  case, relatively speaking. Therefore, efficient grain growth is delayed and the conditions for nucleation stay favourable for longer, resulting in a higher abundance of \textit{J10}  grains. 

An important motivation behind this work was to study uncertainties in the predicted mass-loss rates due to the optical data, since the results of wind model grids are used as input for stellar evolution calculations. The small but systematic differences in mass-loss rates between the \textit{RM}  and \textit{J10}  models reflect a remarkable trend, as compared to the results of \citet{eriksson2023}. These authors investigated the effects dust opacities that are  dependent on the grain size, in contrast to the small particle limit approximation; however, they found no systematic differences in mass-loss rates for these two cases (in contrast to significant differences in grain sizes and photometric properties). Both the present work and the study by \citet{eriksson2023} consider basic assumptions on dust opacities, but the resulting effects on mass loss are qualitatively different. Furthermore, in the models discussed by \citet{eriksson2023}, one of the two alternatives (i.e. the small particle limit) is less general and less consistent, in view of the resulting grain sizes. In the present work, there is no obvious a priori preference for one of the laboratory data sets. Therefore, we investigated various observable properties of the model to see whether one choice gives better agreement with observations than the other. 

Detailed a posteriori radiative transfer calculations were made for a representative selection of the models (46 models in total for each grid), including mainly wind models, along with some models with episodic behaviour and some without wind. The resulting photometric properties show several differences between the two grids of models. These differences mainly concern models with a high C/O ratio, as they produce more dust and the effect of dust opacities becomes stronger as a result. The optical properties of the individual grains and the total amount of dust contribute to higher absorption in the \textit{J10}  models, explaining the trend towards fainter NIR magnitudes. 

Comparisons to observational data have been undertaken, including mass-loss rate versus wind velocity plots and several colour-colour diagrams. 
Both grids show a generally good agreement, especially for models with a carbon excess of 8.5. However, the observed ranges in the CCDs are not fully covered by the synthetic colours of the investigated sub-sample of models, while some models fall outside the observed ranges. Preliminary tests indicate that a better coverage may be achieved by modifying assumptions that affect grain sizes, which should be investigated in a future paper. Based on the current results, there is no clear preference for any one of the data sets when comparing to NIR CCDs.

This work shows that using two different types of commonly applied data for optical properties of amC dust typically results in  a difference in mass-loss rates predicted by DARWIN models by  a factor of 2. The values from earlier C-star model grids (\citealt{mattsson2010,eriksson2014,eriksson2023,bladh2019a}), based on \textit{RM}  opacity data, could possibly be increased within that range, judging from similar models using opacity data by \textit{J10}.

Finally, we note that the current generation of DARWIN models uses a number of approximations as a trade-off for computing efficiency to allow for the creation of large model grids. This concerns both assumptions about micro-physical processes (e.g. position coupling of gas and dust) and overall morphology (spherical symmetry). Taking into account drift between gas and dust or the effects of 3D dynamics (convection, non-radial pulsations, and clumpy atmosphere and wind structures) is beyond the scope of this paper, but should be considered in future studies.

\section*{Data availability}

The appendix of this paper can be found on Zenodo at
\href{https://doi.org/10.5281/zenodo.14904461}{https://doi.org/10.5281/zenodo.14904461}

\bigskip

\begin{acknowledgements}
      We would like to thank Bernhard Aringer for valuable comments on the manuscript.
      This work is part of a project that has received funding from the European Research Council (ERC) under the European Union’s Horizon 2020 research and innovation programme (Grant agreement No. 883867, project EXWINGS), and the Swedish Research Council (Vetenskapsrådet, grant number 2019-04059).
      The computations were enabled by resources provided by the National Academic Infrastructure for Supercomputing in Sweden (NAISS) and the Swedish National Infrastructure for Computing (SNIC) at UPPMAX partially funded by the Swedish Research Council through grant agreements no. 2022-06725 and no. 2018-05973.
      This research has made use of the Spanish Virtual Observatory (https://svo.cab.inta-csic.es) project funded by MCIN/AEI/10.13039/501100011033/ through grant PID2020-112949GB-I00.
\end{acknowledgements}


\bibliographystyle{aa} % style aa.bst
\bibliography{references}

\end{document}